\documentclass[preprint,prl,showpacs,preprintnumbers,amsmath,amssymb]{revtex4}

\usepackage{graphicx}
\usepackage{dcolumn}
\usepackage{bm}

\begin{document}


\title{Image reconstruction without prior information}

\author{K.~S.~Cover (Keith@kscover.ca)}
\affiliation{MEG Centre, VU University Medical Centre, 
             Amsterdam, The Netherlands}

\date{\today}

\begin{abstract}
A novel framework for designing 
image reconstruction algorithms for 
linear forward problems is proposed.  The framework is based on
the novel concept of conserving the information 
in the data during image reconstruction  
rather than supplementing it with prior information.    
The framework offers an explanation as to why the popular reconstruction
algorithms for MRI, CT and convolution are generally
expressible as left invertible matrices.  Also, the framework can 
be used to improve linear deconvolution and tackle such stubborn linear inverse 
problems as the Laplace transform. 
\end{abstract}

\pacs{42.30.Wb 87.57.Nk 87.57.Gg}
\maketitle

The widely used design framework for image reconstruction algorithms of 
linear forward problems is dependent on prior information
\cite{Sabatier2000, Twomey1977, Wiener1949, Park1990, Bert1998, Menke1984, Parker1994, Tara1987, Liang1992}.
Prior information is usually considered
any information not provided by the original data 
(Prior information is defined in more detail below).
But the choice of which prior information to use 
can be a difficult task as different choices can result
in algorithms which generate quite different images
for the same data.  The wide variety of reconstruction 
algorithms of the Laplace transform proposed in the 
literature provides ample evidence
of this situation \cite{Istr1999}.  The complications caused by 
requiring prior information are acknowledged in the 
literature but there is no widely accepted framework
for image reconstruction that does not use prior information. 

To assist in the design of reconstruction algorithms I propose a framework which 
does not use prior information. The framework is free of prior
information, because, as will be proved, 
the algorithms produced by the framework are expressible
as left invertible matrices \cite{Cover2001A,Cover2001B,Cover2003}. 
The choice of which particular 
left invertible matrix to use for the reconstruction 
will depend on the desired focus and noise 
of the reconstructed image and 
the information available in the data via the forward problem.

The required proof will be accomplished 
by showing that for any reconstruction algorithm 
expressible as a left invertible matrix, 
the \( \chi^2 \) fit of any model in model space to the original data 
will be identical to the \( \chi^2 \) fit of the same model 
to the reconstructed image. This property will be referred to as data
conservation.  Thus, left invertible matrices can be thought of as
focusing the data rather than enhancing or altering it
and can be considered digital lenses.

To explain the proposed image reconstruction framework
it is useful to consider as an example a familiar
image reconstruction problem, MRI.  
Conceptually, an MRI medical scanner can be 
thought of as the sole eye witness to a crime. 
After catching a brief glimpse of the suspect, 
it must reconstruct the image by
examining a book of mug shots 
(a model space of composed all 2D integrable functions) and picking
out all mug shots that looked like the perpetrator (any model
which fits the data).  
In the current generally accepted 
framework for the design of
image reconstruction algorithms, the 
scanner's reconstruction algorithm
then picks the one mug shot (or at most a few),
thought to be the ``best'' suspect, based on prior information, 
while discarding all the other pictures that look like the perpetrator.
Thus the question commonly heard when designing a new reconstruction algorithm
is what prior information to use.  It is an important question as a different choice
in prior information can results in a very different reconstructed image
from the same data.

The novel framework described herein 
is based on the concept of reconstructing a single image which is 
comparable to a composite drawing, similar to those routinely issued by police, 
of all the mug shots that look like the suspect.   
I will show no information about the perpetrator 
is lost in the image reconstruction and equally importantly, 
no information is added.
I go on to so show that the
2D FFT reconstruction algorithm used in all, or virtually all, MRI
scanners \cite{Edelstein1980, Chen1989, Park1990, Bert1998}
falls within the proposed framework and does not use prior information.
This is in spite of the fact that the 2D FFT reconstruction algorithm
is usually presented in the literature as using prior information
because prior information is commonly used in its derivation.

For a matrix, \( {\bf A} \), to be considered left invertible, the only
property it must have is that there exists a matrix \( {\bf A}^{-1} \)
that satisfies the equation
\begin{equation}
{\bf I }  = {\bf A}^{-1} \; {\bf A} \label{LeftInvertibleDefined}
\end{equation} 
where \( {\bf I} \) is the identity matrix (+++ find reference other than internet).  
Thus for \( {\bf A} \) to be left invertible it must have the same
number or greater number of rows as columns.   This is important in 
image reconstruction because the reconstructed image
may have many more points then the data. 
An example of image reconstruction 
where there are more values in the image than the 
original data zero filling in MRI reconstruction.

Left invertible matrices should not be confused with 
generalized inverses \cite{BenI1974}.  While the generalized
inverses used in reconstruction algorithms are also left
invertible matrices, left invertible matrices provide a
far larger set of matrices from which to select an
matrix with the desired focus and noise characteristics.
Generalized inverses will be discussed in more detail below.

To reconstruct the composite image 
generated by the ``sole eye witness''
we first need a mathematical expression of the 
scanner's measurement properties.
An equation of this type is called a forward problem.  The
general form of the linear forward problem is 
\begin{equation}
{\bf d} = \int_a^b \!  m^O(y)\, {\bf G} (y)\, dy + {\bf e}
\label{IntroForward}
\end{equation}
where the data vector, \( \bf d\), and the data kernel, \( {\bf G} (y) \) 
are assumed known.  The scanned object, \( m^O(y) \),
is represented as a model in the model space and is the model of
which we want to reconstruct an image. 
The noise, \( \bf e \), is assumed to be additive,
stationary and have a mean of zero.  The noise statistics are
characterized by a covariance matrix, \( \bf C^d \).  
The bounds of integration, \( a \) and \( b \), are dependent on the 
particular forward problem.

The MRI scanner measures (sees) a Fourier transform of the
object being scanned (the perpetrator).  
For the 1D Fourier transform the forward problem has the form
\begin{equation}
d_n = \int_0^1 \!  m^O(f)\, e^{2\pi i f n/N}   df + e_n
\label{IntroForwardFFT}
\end{equation}
where the index of the data points, \( n \), has a range of 0 to \( N-1\)
where \( N \) is the number of data points.
Both Eqs. (\ref{IntroForward}) and (\ref{IntroForwardFFT}) are
given in 1D for simplicity, but they can be easily generalized
to the 2D FFT used in MRI reconstruction.

The forward problem in Eq. \ref{IntroForward}, and 
in particular, the Fourier transform forward problem in Eq. \ref{IntroForwardFFT}
are ill posed problems.  This is because the finite number of data points
and the noise in the data both allow many different models in 
model space to fit the data reasonably well \cite{Olden1975}. 
It is the ill posed nature of forward problem which presented such
a challenge to image reconstruction for many decades.

The currently used, prior information framework for image reconstruction 
needs some measure of how well a model 
in model space fits the data
so that, conceptually at least, a subspace of models
that fit the data can be constructed.
In our analogy, after the MRI scanner has made its measurements, 
it can be thought of as viewing
all mug shots in the book (all 2D functions)
and choosing all the ones that look like the perpetrator.
The most commonly used measures of whether a particular 
model in model space
fits the data are the \( \chi^2 \) measure and the 
least squares measure \cite{Twomey1977, Menke1984, Tara1987}.
The least squares measure is equal to the \( \chi^2 \) measure
if the noise covariance matrix
is set equal to the identity matrix. 
Therefore, only the \( \chi^2 \) measure
will be considered for the rest of this letter.
The \( \chi^2 \) measure is calculated by 
\begin{equation}
\chi^2 =  ({\bf d} -{\bf d}^m)^T  {\bf C^d}^{-1} ({\bf d} -{\bf d}^m)
\label{chi2Data}
\end{equation}
where \( {\bf d}^m \) is the data predicted by each model in
model space and is calculated by 
\begin{equation}
{\bf d}^m = \int_a^b m(y)\, {\bf G}(y) \, dy.
\label{model2Data}
\end{equation}

In the proposed framework the reconstruction algorithm 
is assumed to be expressible as a 
matrix multiplication.  Thus, the form of the image reconstruction algorithm is 
\begin{equation}
{\bf h} = {\bf A \; d}
\label{ImageReconMatrixToData}
\end{equation}
where \( \bf A \) is the image reconstruction matrix and
\( \bf h \) is the reconstructed image.
For calculation purposes, the image is assumed discretized into
\( M \) points where the discretization is sufficiently finely 
discretized to approximate a continuous function.  The the
reconstruction matrix, \( A \), has \( M \) rows and \( N \) columns. 
It follows that the covariance of the noise in the image, \( \bf C^h \), is calculated by
\begin{equation}
{\bf C^h} =   {\bf A C^d} {\bf A}^T.
\label{ImageCovariance}
\end{equation}
Also, the image kernel (resolution kernel), \( {\bf R}(y) \), is calculated by   
\begin{equation}
{\bf R}(y) = {\bf A} \; {\bf G}(y).
\label{ImageReconMatrixToDF}
\end{equation}

In the prior information framework 
of linear inverse theory, the reconstructed image is
thought of as the best image (model) that fits the data
where what is best is decided with the aid of prior information,
information in addition to the data.
The proposed framework breaks with this practice 
by considering the reconstructed image to be a reformulated forward problem.
Further, I will conjecture that it is the current practice
of radiologists when interpret MRI images, they are interpreting
as forward problems rather than models.

Examination of Eq. (\ref{IntroForward}) 
shows that a forward problem is composed of 
three parts: data, data noise statistics and a data kernel.
Examination of Eqs. (\ref{ImageReconMatrixToData}),
(\ref{ImageCovariance}) and (\ref{ImageReconMatrixToDF})
shows that the image also has these three components. 
Thus an image reconstructed by an algorithm representable
as a matrix multiplication also has the
mathematical form of a forward problem.    
In addition, the original and reformulated image  
forward problems are defined on the same model space.

As mentioned above, when \( \bf A \) is left invertible,
a property that the original and image forward
problems share  
is that any model in model space will have the identical \( \chi^2 \)
measure for the original and image forward problem.  This property
of data conservation
means that the original and image forward problems will 
have exactly the same information about the scanned object.
Data conservation is one of the two key properties
of the proposed framework.  The second is data focusing and
will be considered below.

The concept of data conservation and prior information are closely linked.
If the original data and the reconstructed image have exactly the same
information about the scanned object then no additional information
could have been added or lost during the image reconstruction.
Thus, either prior information was not used in the reconstruction,
or if it was, it had no impact on the reconstruction.
If prior information did have an impact on the reconstruction
then data conservation would be violated.
Thus, prior information can be defined as the absence
of data conservation.

The proof of data conservation is as follows:
The \( \chi^2 \) measure for the image forward problem is 
\begin{equation}
\chi^2 = ({\bf h} - {\bf h}^m)^T {\bf C^h}^{-1} ({\bf h} - {\bf h}^m)
\label{chi2Image}
\end{equation}
where \( {\bf h}^m \) is calculated by substituting Eq.
(\ref{model2Data}) into Eq. (\ref{ImageReconMatrixToData}). 
If data conservation holds when \( \bf A \)
is left invertible then Eq. (\ref{chi2Image}) will equal to Eq. (\ref{chi2Data}).
Eq. (\ref{chi2Image}) simplifies to (\ref{chi2Data})
when Eqs. (\ref{ImageReconMatrixToData}), (\ref{ImageCovariance}) and 
\begin{equation}
{{\bf C^h}^{-1}} = {\bf A^T}^{-1} {\bf C^d}^{-1} {\bf A}^{-1}.
\label{ImageCovarianceInverse}
\end{equation}
are substituted into (\ref{chi2Image}).
Eq. (\ref{ImageCovarianceInverse}) is used to calculate
\( {\bf C^d}^{-1} \) from \( {\bf C^h}^{-1} \) and follows
from Rao \cite{Rao1973}.
Thus, the \( \chi^2 \) measure of fit
for any model in model space 
will be identical for the original data and image
forward problems if the reconstruction algorithm is representable
as a left invertible matrix.  Therefore, data conservation
holds if the reconstruction matrix is left invertible.

In addition to data conservation, the proposed framework
uses the familiar concept of focused data.  
The parallels between focusing using a matrix
and focusing with a glass lens have been understood
for many years \cite{Fell1952}.  
Focused data is used in Wiener deconvolution \cite{Wiener1949},
the 2D FFT and filtered back project. 
The left invertible matrices used in these algorithms generates 
data functions in the image kernel that are  
much more similar to Dirac delta function than in the
original data kernel, thus focusing the data (Eq. \ref{ImageReconMatrixToDF}).  
However, while choosing a matrix \( \bf A \)
which improves the focus of the image kernel, 
the corresponding increase in noise of the image 
must not be allowed to get too large (Eq. \ref{ImageCovariance}).  This
choice of focus and trade off between focus and noise is an inherent part of the design
in any linear reconstruction algorithm.  It is part of 
the reason why a window is routinely used in MRI reconstruction
and a filter used in filtered back projection.
It is also explicitly dealt with in Wiener deconvolution. 
Backus and Gilbert \cite{Backus1970} also use the concept
of focus data and the tradeoff between resolution and
noise but do not use the concept of data conservation.


The prior information framework for the design of image reconstruction
algorithms is based on the requirement of fitting the data.  
This requirement demands that any reconstruction algorithm produced
by the framework must generate an image that, when inserted back into
the forward problem (Eq. \ref{IntroForward}), reproduces the data
exactly or at least to within the noise.  If the reconstruction
algorithm is linear (Eq. \ref{ImageReconMatrixToData}) the exact fitting
constraint of the prior information framework leads to the requirement
on the reconstruction matrix, \( \bf A \), that
\begin{equation}
{\bf I }  = {\bf G} \; {\bf A} \label{GeneralizedInverseDefined}
\end{equation}
where \( \bf I \) is the identity matrix with dimensions \( N \times N \)
and \( G \) is the discretized form of the forward problem (Eq. \ref{IntroForward})
and has \( M \) rows and \( N \) columns.

It is illuminating to compare Eq. \ref{GeneralizedInverseDefined} with Eq. \ref{LeftInvertibleDefined}
as they provide the primary restrictions on the reconstruction matrices in each of their
respective frameworks.  The restriction of left invertibility, which results in data conservation,
is much less restrictive than that of generalized invertibility.  This is the case
because in Eq. \ref{LeftInvertibleDefined} the matrix \( {\bf A}^{-1} \) can be
any matrix while in Eq. \ref{GeneralizedInverseDefined} it must be \( G \).
Experience has shown, the much wider choice in reconstruction matrix
provided by data conservation in many cases
permits the choice of a reconstruction matrix with far superior focus and noise characteristics. 
Examples of reconstruction algorithms which fit into this framework
are the standard MRI and CT reconstruction algorithms.

Comparison of equations Eq. \ref{GeneralizedInverseDefined} with Eq. \ref{LeftInvertibleDefined}
also shows that generalized inverses are left invertible and are thus data conserving. 
But Eq. \ref{GeneralizedInverseDefined} applies only to reconstruction algorithms which produce
exact fits to the data.  But not all reconstruction algorithms produce exact fits.
Some forms of regularization used in the prior information 
framework relax the fitting constraint slightly
by introducing a misfit parameter, such as \( \chi^2 \).  
In this case Eq. \ref{GeneralizedInverseDefined} is only satisfied approximately
and, as a consequence, data conversation only applied approximately.  

Frequently, in the prior information framework, nonlinear reconstructions algorithms
are used.  The left invertible matrix is a linear reconstruction algorithm thus
data conservation will need to be proved for nonlinear reconstruction algorithms
case by case.

It is useful to consider three highly popular reconstruction algorithms
in terms of the data conservation and prior information frameworks: 
1) the windowed 2D FFT used in MRI reconstruction, 
2) the filtered back projection used in CT reconstruction and 
3) Wiener deconvolution for linear deconvolution.
The 2D FFT without the window, back projection without the filter and
Wiener deconvolution all fit into the prior information framework
as reconstructed image fits the original data.  As is well
known, all three algorithms can be derived using 
the least squares fit to the data.  They also
fit into the data conversation framework as they are left
invertible algorithms.  In all three cases the image kernel
is much better focused then the data kernel.

In routine use the 2D FFT and back projection are
modified by a window and filter respectively \cite{Harris1978, Farq1998}
to tailor the focus and noise performance of the reconstruction
algorithms to the needs of their daily applications.  When the window
or filter is introduced Eq. \ref{GeneralizedInverseDefined} is
no longer satisfied.  For example, in the case of the windowed 2D FFT
the left hand side will equal the coefficients of the window,
which is very different from the identity matrix.    
Thus the reconstruction algorithms currently used
do not fit into the prior information framework.  But,
as long as all coefficients of the windows and filters
are nonzero, the algorithms are left invertible, and thus
still data conserving.

It is conjectured above radiologist routinely
interpret MRI images as forward problems.  
To do so they would need to consider, not
only the data in the image, but image noise and
image kernel functions.  When radiologists look at
an MRI image they take note of the noise and
resolution of the image when making their 
interpretations.  For most clinical MRI
images the signal to noise is reasonably
good and the resolution has a width below
1mm due to the 
image kernel functions being similar to Dirac delta functions.  
Most models which fit poorly to the image
are quickly and easily ruled out
because of the nature of the \( \chi^2 \) fit.  
If a model is a bad fit
to one part of an image it must
also be a bad fit to the image as a whole
because the value of the \( \chi^2 \) cannot 
be made smaller by the rest of the image. 
Thus radiologists are taking advantage of 
data focusing and data conservation        
to effectively rule in and out models in their head
treating the image as a forward problem.  Thus
they are treating the reconstructed image a
composite of possible brains than as one particular brain.

There are many ways to generate a left invertible matrix
with a good focus.  For the inverse Fourier transform, 
linear convolution and the back projection problem, 
the least squares fit to the data usually results in a 
left invertible matrix with reasonable
good focus.  I conjecture the least squares fit works so well
on these three problems because the least squares fits
produces the same image as Backus and Gilbert's
Dirichlet focusing criterion \cite{Backus1970}.
The Dirichlet criterion appears to produce good focus
for images that have spatial invariant image kernels.
This is true for the Fourier transform, Wiener deconvolution
and approximately true near the center of
image in filtered back projection.

But the least squares fit does not always generate a 
left invertible matrix with a good focus.  Applying
the least square fit to the Laplace transform yields
a reconstruction matrix which is left invertible,
provided enough rows are used, but with a terrible focus.  
The proposed framework has been
applied to the Laplace transform to generate
a left invertible matrix with a good focus and 
reasonable noise characteristic \cite{Cover2001B}.
One guideline which is useful in the design of a left
invertible matrix is that each point in the image
has the same same standard deviation for the noise.
As each point in the image corresponds to a row
in the reconstruction matrix, each row
can be tailored individually to get the
best combined of resolution and noise.  
For particular problems, such as the inverse Laplace transform,
it maybe be necessary to design several reconstruction matrices
each which produces a different noise in the image.  The various
reconstruction algorithms can then be applied a few typical
data sets to determined the desired tradeoff between focus and noise.

The proposed framework could easily be applied to linear deconvolution
by designing a linear deconvolution operator which is
left invertible and which has different
focusing properties than Wiener deconvolution, for example
with less side lobes.

Many researchers believe that solving any linear inverse problem,
including image reconstruction, requires information
in addition to that available in the data.  
However, the proposed framework shows there exist 
image reconstruction algorithms that neither
add nor take away information to the data.  
Instead, they only focus the data making it
easier to interpret and manage.
After data conserving reconstruction, the focused data 
can be used for interpretation, 
model fitting and most other 
processes intended for the original data.
It was shown the standard reconstruction algorithms for
MRI, CT and linear deconvolution are already data conserving.
Thus, image reconstruction algorithms which are both 
data conserving and data focusing already play
a large role in modern data analysis and may have
an even larger one in the future.

\begin{acknowledgments}
Research was supported by D.~W.~Paty,  
D.~K.~B.~Li and the MS/MRI
Research Group of the University of British Columbia. 
I thank A.~L.~MacKay, B.W.~van~Dijk, J.J.M.~Zwanenburg, J.P.A.~Kuijer and H.~Vrenken 
for comments.  
\end{acknowledgments}


\end{document}